\documentclass[prb,superscriptaddress,preprintnumbers,a4paper,amsmath,amssymb,showpacs,floatfix]{revtex4-1}
\usepackage{graphicx}
\usepackage{dcolumn}
\usepackage{bm}

\usepackage{times}
\usepackage{color}
\newcommand{\fig}[1]{figure~\ref{#1}}

\newlength{\figwidth}
\newcommand{\tabl}[1]{table~\ref{#1}}

\newlength{\hw}
\newlength{\vvp}
\newlength{\minusspace}
\newcommand{\msp}{\hspace{\minusspace}}
\newlength{\zerospace}
\newcommand{\degree}{$^\circ$}
\newcommand{\mub}{$\mu_{\mathrm B}$}
\newcommand{\sinth}{\ensuremath{\sin\theta/\lambda}}
\newcommand{\inA}{\AA$^{-1}$}
\newcommand{\smgd}{Sm$_{.976}$Gd$_{.024}$Al$_2$}
\newcommand{\smgdx}{Sm$_{1-x}$Gd$_{x}$Al$_2$}
\newcommand{\smal}{SmAl$_2$}
\newcommand{\smion}{Sm$^{3+}$}
\newcommand{\eith}{\ensuremath{\frac18}}
\newcommand{\hf}{\ensuremath{\frac12}}
\newcommand{\jzero}{\ensuremath{\langle j_0\rangle}}
\newcommand{\jtwo}{\ensuremath{\langle j_2\rangle}}
\newcommand{\jzerok}{\ensuremath{\langle j_0(k)\rangle}}
\newcommand{\jtwok}{\ensuremath{\langle j_2(k)\rangle}}
\newcommand{\Gsm}{\ensuremath{G_{\textrm{\scriptsize Re}}}}
\newcommand{\Gal}{\ensuremath{G_{\textrm{\scriptsize Al}}}}
\newcommand{\Tsm}{\ensuremath{T_{\textrm{\scriptsize Re}}}}
\newcommand{\Tal}{\ensuremath{T_{\textrm{\scriptsize Al}}}}
\newcommand{\bsm}{\ensuremath{b_{\textrm{\scriptsize Re}}}}
\newcommand{\bal}{\ensuremath{b_{\textrm{\scriptsize Al}}}}
\newcommand{\Bsm}{\ensuremath{B_{\textrm{\scriptsize Re}}}}
\newcommand{\Bal}{\ensuremath{B_{\textrm{\scriptsize Al}}}}
\newcommand{\Tcomp}{\ensuremath{T_{\textrm{\scriptsize comp}}}}
\newcommand{\Nobs}{\ensuremath{N_{\textrm{\scriptsize obs}}}}
\newcommand{\Fobs}{\ensuremath{F_{\textrm{\scriptsize obs}}}}
\newcommand{\Fcalc}{\ensuremath{F_{\textrm{\scriptsize calc}}}}
\begin{document}
\title{Polarized single crystal neutron diffraction study of the zero-magnetization ferromagnet Sm$_{1-x}$Gd$_x$Al$_2$ (x = 0.024)}
\author{T. Chatterji}
\email[Email of corresponding author: ]{chatterji@ill.fr}
\affiliation{Institut Laue-Langevin, 71 Avenue des Martyrs, 38000 Grenoble, France}
\author{  A. Stunault}
\affiliation{Institut Laue-Langevin, 71 Avenue des Martyrs, 38000 Grenoble, France}
\author{ P.J. Brown}
\affiliation{Institut Laue-Langevin, 71 Avenue des Martyrs, 38000 Grenoble, France}


\date{\today}
\begin{abstract}
We have determined the temperature evolution of the spin and orbital moments in the  zero magnetization ferromagnet Sm$_{1-x}$Gd$_x$Al$_2$ (x = 0.024) by combining polarized and unpolarized single crystal neutron diffraction data. The sensitivity of the polarized neutron technique has allowed the moment values to be determined with a precision of $\approx 0.1$~\mub. Our results clearly demonstrate that, when magnetised by a field of 8T, the spin and orbital moments in Sm$_{1-x}$Gd$_x$Al$_2$ are oppositely directed so that the net magnetization is very small. Below 60 K  the contributions from spin and orbital motions are both about 2\mub\ with that due to orbital motion  being slightly larger than that due to spin. Between 60 and 65 K the contributions of each to the magnetization  fall
rapidly and change sign at \Tcomp\ $\approx 67$K above which the aligned moments recover but with the orbital magnetization still slightly higher than the spin one. These results imply that above \Tcomp\ the small resultant magnetization of the \smion\ ion is 
oppositely directed to the magnetizing field. It is suggested that this anomaly is due to polarization of conduction electron spin associated with the doping Gd$^{3+}$ ions.
 \end{abstract}
 \maketitle 

\section {Introduction}
   Strong spin-orbit interaction \cite{vanvleck32,zeiger73,stevens97} is known to lead to interesting physics in, for example,  topological insulators \cite{hasan10,kane05a,kane05b}, thermoelectric materials \cite{wan10,boukai08,ohta07,heremans08,zhao14}, Dzyaloshinskii-Moriya weak ferromagnets \cite{turov65} etc. Amongst these the rare-earth ion Sm$^{3+}$ presents a unique case since the ion has spin and orbital magnetic moments, both of about 4 $\mu_B$, which are strongly coupled antiparallel to one another so that they almost cancel out. 
   In most rare-earth elements, the ground $J$ multiplet is sufficiently separated from other multiplets that both the spin and orbital moments can be expressed in terms of the same operator of the total angular momentum. This implies that they have the same temperature dependence. However for \smion\ a few low level multiplets 
with different  $J$ values lie close enough in energy to the ground multiplet to mix with it. Due to this mixing the spin and orbital moments in \smion\ depend upon distinct operators. The degree of admixture of  close multiplets, and its temperature variation in the solid, may lead to exact cancellation of the spin and orbital moments in a narrow range of temperature; giving rise to the so called zero-magnetization ferromagnet (ZMF).  Studies of the temperature dependence of magnetization in  SmAl$_2$ suggest that in the pure compound the admixture of multiplets is not enough to cause complete cancellation, however doping SmAl$_2$ with Gd$^{3+}$  which has a large spin-only moment, does lead to  a zero-magnetization ferromagnet Sm$_{2-x}$Gd$_x$Al$_2$ \cite{adachi99}.
 For $x=0.0185$ the compensation temperature \Tcomp\ $\approx 80$K with  a ferromagnetic transition temperature $T_C \approx 120$ K. 
 
   \vspace{2ex}

   The unique properties of zero-magnetization ferromagnets are well suited for use in devices processing the spin of  charged particles. ZMF materials, despite their uniform spin polarization, do not generate stray magnetic fields which  perturb the motion of charged particles and so can be used in spin electronics manipulating both electric current and spin polarization.
\vspace{2ex}

   The behaviour with varying temperature  of the oppositely oriented spin and orbital moments of \smion\ in a
solid state environment has been studied in the zero magnetization ferromagnet \smgdx\ $x=0.18$ using several different techniques \cite{adachi01,taylor02,qiao04,pratt06}. It is not easy to determine the orbital and spin moments separately. Neutron diffraction cannot separate these moments directly, but they do have distinct cross-sections for scattering of elliptically polarized  X-rays. Results, using this technique, show a distinct cross-over of spin and orbital moments at \Tcomp\ $\approx 80$~K \cite{taylor02}. Qiao et. al. \cite{qiao04} used X-ray magnetic circular dichroism to determine the temperature dependence of the spin and orbital contributions to the moments of the Sm and Gd ions separately. They conclude that conduction electrons contribute almost as much to the magnetization as the spin of \smion\ with the same temperature variation.
In complement to these studies of spin and orbital moments. helicity switching Compton scattering \cite{adachi01} and specific heat measurements \cite{taylor02} have  been used to prove that ferromagnetic order persists through the whole of the compensation region.  $\mu$SR investigations \cite{pratt06} have also been carried out  on these ZMF Sm$_{1-x}$Gd$_x$Al$_2$ materials. Gotsis et al. \cite{gotsis03} have calculated the electronic structure and magnetic properties of ferromagnetic SmAl$_2$ using the local spin-density approximation LSDA + U, they found that this method can give a physically meaningful description of spin-orbit compensation in Gd doped SmAl$_2$.

  \vspace{2ex}

Although, as noted above, neutron diffraction cannot separate the spin and orbital contributions to the magnetic
moments directly, the cross-section for magnetic neutron scattering by the \smion\ ion can be modelled  
in terms of these two parameters using the dipole approximation \cite{marshall:71}.  In this approximation the 
amplitude of magnetic scattering by the ion may be expressed as a function of the scattering vector k as
\begin{figure}[h]
\begin{center}
\resizebox {0.5\textwidth}{!}{\includegraphics{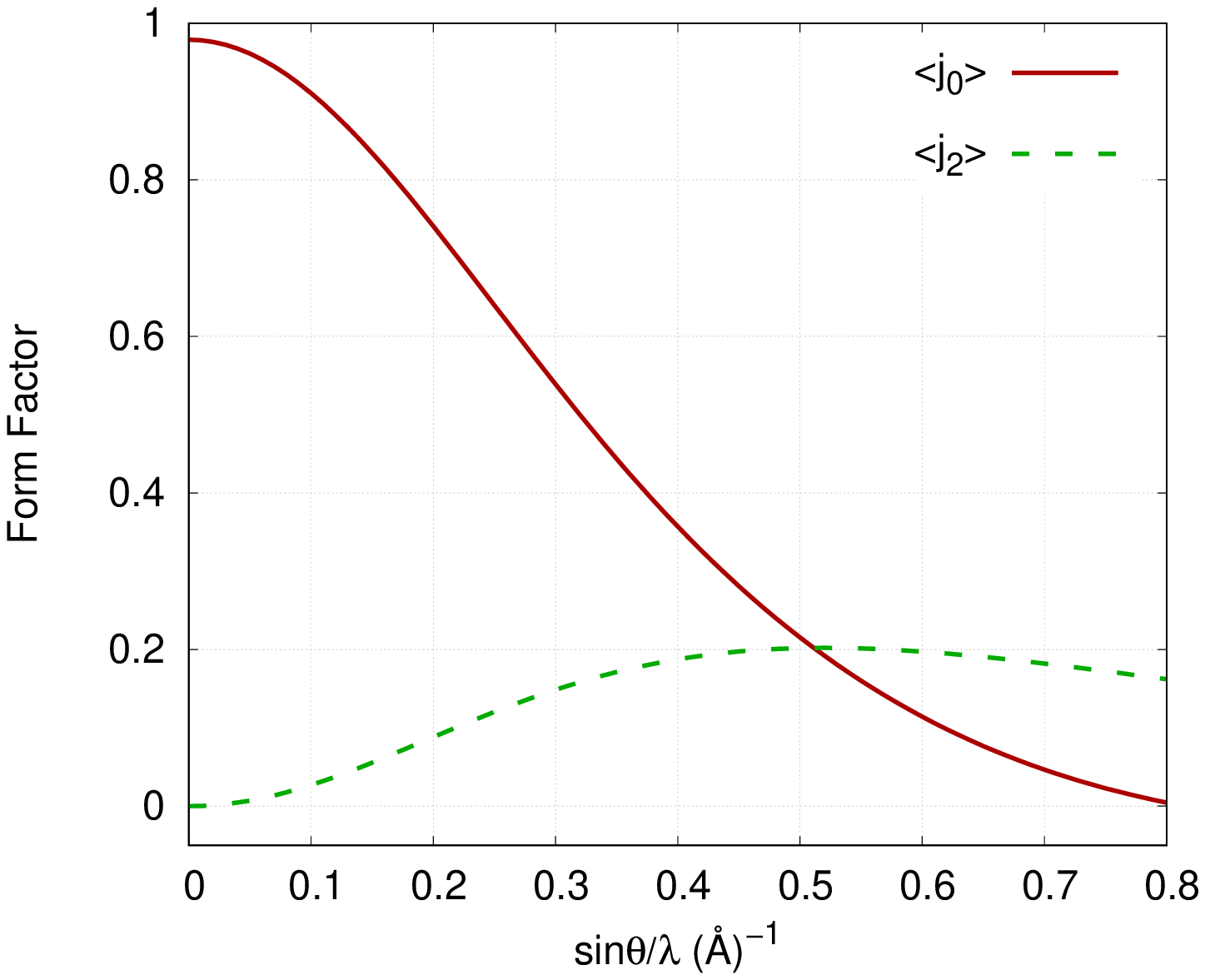}}
\caption{Form factors \cite{blume:62} \jzero\ and \jtwo\  for \smion\ }
\label{smff}
\end{center}
\end{figure}
\begin{equation} \label{dipole}
f(k)=\jzerok{\bf\hat S} + \hf(\jzerok + \jtwok){\bf\hat L}
\end{equation}
Here ${\bf\hat S}$ and ${\bf\hat L}$ are the mean spin and orbital quantum numbers for the ion and the form-factors
${\langle j_l(k)\rangle}$ are calculated from the radial distribution $U^2(r)$ of the 4f electrons in the ion using
\begin{equation}
\left\langle j_l(k)\right\rangle= \int U^2(r)j_l(k r)4\pi r^2\,dr
\end{equation}
in which the $j_l(kr)$ are spherical Bessel functions.
The variation with $k \propto \sin\theta/\lambda$ of \jzero\ and  \jtwo\ for \smion\
is shown in \fig{smff} 
from which it can be seen that they vary very differently between $k=0$ and $0.8 $~\inA. We have exploited this
difference to determine  the temperature variations of the spin and orbital contributions to the magnetic moment
of \smion\ in \smgd\  separately. 
Although the total magnetization of the ZMF \smgd\ is rather small these separate components can be determined with high precision using 
polarised neutron
diffraction because the polarised neutron intensity asymmetry depends on the ratio between the magnetic and nuclear
structure factors rather than on the sum of their squares as is the case for unpolarised neutrons.
To determine magnetic form factors from polarised neutron asymmetries the nuclear structure factors
must be known to the required precision and in particular extinction effects, which invalidate the proportionality 
between the scattered intensity and the square of the structure factor, must be modelled accurately.
  \vspace{2ex}

 Neutron diffraction 
studies of Gd and Sm compounds are difficult because of the high absorption cross-sections of both these elements for thermal neutrons ($> 10^4$ barns) \cite{lynn:90}. This difficulty can be greatly reduced by using "hot" neutrons available from the ILL hot source. With neutron energies of $\approx$ 300 mev ($\lambda=0.5$~\AA)
the absorption cross-section is reduced b a factor of about 40 leading to a linear absorption coefficient  
less than 1~mm$^{-1}$ for \smal. The polarised and unpolarised neutron diffraction measurements described
here were made possible by using these shorter wavelengths. SmAl$_2$ has the C14 cubic Laves phase structure, space group Fd3m. Both the Sm and Al atoms
occupy special positions: Sm  8a (\eith,\eith,\eith), Al 16d (\hf,\hf,\hf).
The only free parameters affecting the nuclear structure factors are the scattering length of the rare-earth site
(SmGd), which is not accurately known at short wavelengths, and the Sm and Al isotropic temperature factors. It may
be noted that at these short wavelengths the contribution to the scattering cross-sections of the imaginary part of the rare earth scattering length ($\approx 0.2$ fm) can be neglected.

\section{Experimental}
Single crystals of  Sm$_{.976}$Gd$_{.024}$Al$_2$ were grown from the melt using the  Bridgeman and also Czochralski methods. 
Neutron diffraction measurements were made on two crystals of nominal composition Sm$_{0.976}$Gd$_{0.024}$Al$_2$  and
approximate volumes 500 mm$^3$ (crystal X1) and 20 mm$^3$ (crystal X2).
\subsection{Unpolarised Neutron intensity measurements}
Integrated intensity measurements were made on the 4-circle hot source diffractometer D9 at the Institut Laue Langevin Grenoble. The sample temperature was controlled  by a 2-stage displex refrigerator. 
The integrated intensities of all accessible reflections  with $\sinth < 0.85$~\inA\ were measured from X1
at temperatures of 30, 62 and 100K using a neutron wavelength 0.51 \AA. The cubic symmetry allowed many equivalents
of each of the independent reflections to be measured. These showed a spread in intensity of up to a factor 2 which
was attributed to the variations in absorption due to the asymmetric shape of the crystal. 
Intensities measured up to \sinth=0.5 \inA\ from the much smaller crystal X2 at 70~K showed, as expected, 
much less divergence between  equivalent reflections. 
\subsection{Polarised Neutron intensity asymmetry}
Polarised neutron measurements were made using the spin polarised diffractometer D3 at ILL which also receives neutrons from the hot source. The crystals were mounted in an asymmetric split-pair cryomagnet and magnetised
with a vertical field of 8T.
The asymmetry in the peak intensities of Bragg reflections for 0.52 \AA\ neutrons polarized parallel and antiparallel 
to the field direction were made on both crystals at temperatures in the range 32 - 105~K. With the large crystal X1 a significant asymmetry was measured in  24 independent reflections with $0.21<\sinth<1.14$ \inA whereas  for
the small crystal X2 only 12 independent reflections, $0.21<\sinth<0.72$ \inA could be measured.

\section {Results}
\subsection{Nuclear structure model}
A linear absorption coefficient $\mu=0.38$~mm$^{=1}$ for $\lambda=0.51$~\AA\ was estimated from the curve of total cross-section vs energy given by Lynn and Seeger\cite{lynn:90}. Transmission factors for all measured reflections 
were calculated using this coefficient and approximate models of the crystal shapes. Applying this correction
led to a marked  reduction in the spread of intensities for the symmetry related reflections from crystal X1.  
 The transmission factors calculated for this crystal  varied between 0.04 and 0.184 whereas for crystal 2 the range was only 0.50 to 0.53.
 
 After correcting for absorption, the mean structure amplitude was calculated  for each independent reflection in the sets   measured 
at each  temperature. These structure amplitudes were used as data in
least squares refinements  in which initially the free parameters were a scale factor, 
the Sm site scattering length \bsm, the Sm and Al isotropic temperature factors 
\Bsm\, \Bal\ and a single extinction parameter (mosaic spread). These initial refinements led to unphysical (negative) values for the mosaic spread parameter, suggesting that the degree of extinction was small. They also indicated a high degree of correlation between the isotropic temperature factors and the other parameters. Further refinements were carried out without extinction giving 
results shown in the first part (A) of table~\ref{strucp}. 
\begin{table}[h]
\caption{Results obtained from least squares refinements of the nuclear structure parameters of \smgd}
\label{strucp}
\begin{center}
(A) All parameters varied\\[.4ex]
\begin{tabular}{lcllllcl}
\hline
T&Crystal&\bsm\ (fm) & \Bsm\ (\AA$^{-2}$)&\Bal\ (\AA$^{-2})$&Scale&\Nobs &R$_{\textrm{\scriptsize{cryst}}}$\footnote
{R$_{\textrm{\scriptsize{cryst}}}=\left(\sum_1^{\Nobs}|\Fobs-\Fcalc|\right)/\left(\sum_1^{\Nobs}|\Fobs|\right)$ where \Fobs\ and \Fcalc\ 
are the absolute values of the observed and calculated structure factors and \Nobs\ the number of observations,}
 (\%)\\
\hline
30&X1& 5.9(3)&  0.42(11)& 0.69(9)&19.7(6)& 55& 7.1\\
62&X1& 6.0(1)&  0.29(5)& 0.44(6) &19.6(5)& 72& 6.3\\
100&X1& 5.2(2)&  0.19(6)& 0.51(6) &19.4(3)& 70& 5.3\\
70&X2& 5.2(3)& -0.2(2)&  0.3(2)  &2.31(8)& 20& 4.5\\
\hline
\vspace{-1ex}
\end{tabular}\\
(B) \bsm\ = 5.65(23) fm; Scale for X1=19.5(4)\\[.4ex]
\begin{tabular} {lclll}
\hline
T&Crystal&\Bsm\ (\AA$^{-2})$&\Bal\ (\AA$^{-2})$&R$_{\textrm{\scriptsize{cryst}}}$ (\%)\\
\hline
30&X1& 0.28(5) & 0.69(5) &7.1\\
62&X1& 0.20(3) & 0.45(3) &5.9\\
100&X1&0.38(3) & 0.52(3) &5.5\\
70&X2& 0.03(20) & 0.11(18) & 4.4\\
\hline
 \end{tabular}
\end{center}
\end{table}%
To reduce the effects of correlation in further refinements the Sm scattering length 
was fixed to the value 5.65~fm obtained from the weighted mean of its 4 values in (A) and the scale
for all X1 data to the mean of the 3 values obtained with \bsm\ fixed. 
The final results for which only the temperature factors were refined are in table~\ref{strucp}(B). Since the amplitudes of thermal vibrations are
not expected to vary much in this temperature range and the differences in the refined $B$ values are hardly significant, the mean values $\Bsm = 0.29(5)$ and $\Bal = 0.51(5)$ were used at all temperatures in subsequent calculations. 

\begin{figure}[t]
\begin{center}
\resizebox{0.75\textwidth}{!}{\includegraphics{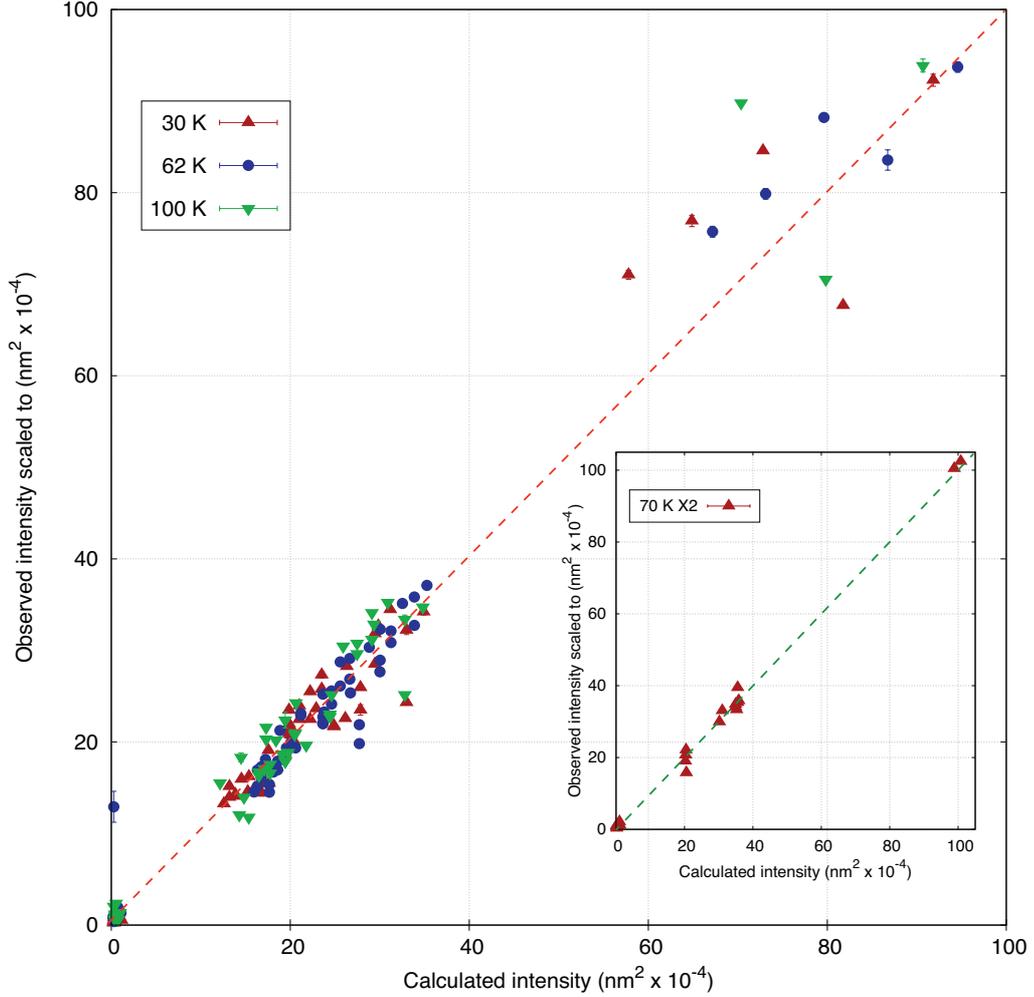}}
\caption{The squares of the structure factors measured for crystal X1 at three temperatures plotted against their values calculated from model B. The inset shows the same plot for reflections measured from the smaller
crystal X2.}
\label{xtest}
\end{center}
\end{figure}

Figure \ref{xtest} shows the squares of the experimental structure factors plotted against the corresponding values
calculated for the model structure B. It can be seen that there is a good linear dependence between the two. The scatter in the high intensity reflections from crystal X1 is probably due to inadequacy in the absorption corrections due to difficulty in accurately describing the crystal shape. Most importantly the good linearity
confirms the absence of any significant extinction allowing the polarized neutron intensities to be analysed 
using a zero extinction model.
\subsection{Spin and Orbital moments from polarised neutron asymmetry }
The polarized neutron intensity asymmetry for a magnetized ferromagnet is defined as 
$A = (I^+ - I^-)/(I^+ + I^-)$ where $I^+$ and $I^-$ are the 
intensities measured with neutrons polarised parallel and antiparallel to the magnetizing field.
In the absence of extinction the asymmetry measured  with polarizing efficiency $P$ for a reflection with scattering vector
inclined at an angle $90-\rho$ to the field is given in terms of the magnetic and nuclear structure factors $F_M$ and $F_N$ by
\begin{equation}
A=\frac{2PF_NF_M\cos^2\rho}{F_N^2+F_M^2q^4}=\frac{Pq\gamma}{1+q^2\gamma^2}\quad\mbox{with}\quad q=\cos^2\rho \quad\mbox{and}\quad\gamma = F_M/F_N
\label{atogam}
\end{equation}
Equation \ref{atogam} was used to calculate the ratio $\gamma$ for all the asymmetries measured in the experiment. 
and these were than used to determine the temperature variation of the aligned magnetization and its form factor.

The magnetic structure factor for \smal\  depends just on the rare earth ion   
$$F_M(k)=\mu_0 f(k)\Gsm\Tsm$$
 where $\mu_0$ is the Sm magetic moment $f(k)$ its magnetic form factor, \Gsm\ its geometric structure factor and  \Tsm\ its isotropic temperature factor. The nuclear structure factor on the other hand depends on both Sm and Al
$$F_N(k)=\bsm\Gsm\Tsm + \bal\Gal\Tal$$  where  \bsm\ and \bal\ are the neutron scattering lengths, giving
\begin{equation} \label{mugam}
\mu_0f(k)=\gamma\left(\bsm+\bal\frac{\Gal\Tal}{\Gsm\Tsm}\right) = \gamma(\bsm + \bal R)
\end{equation} 

Since the purpose of 
the experiment is to determine the magnetic form-factor for Sm one must consider to what extent uncertainties
in the  parameters of the model may affect the result. For \smal\ these are in just \bsm\ and the ratio $\Tal/\Tsm$.
The ratio $\Tal/\Tsm$ can be written as $\exp -k^2(\Bal - \Bsm)$ so the uncertainty in the temperature factors gives
an extra uncertainty in $\gamma$ proportional to $Rk^2\sqrt((\Delta\Bal)^2 + (\Delta\bsm)^2)$ where the $\Delta$s represent the 
estimated standard deviation (esd) in the following parameter.
 
 Including all contributions
\begin{equation} \label{errgam}
\Delta(\mu_0f(k))=\sqrt{(\Delta\gamma(\bsm + \bal R))^2 + (\gamma\Delta\bsm)^2 + 
(\bal Rk^2[(\Delta\Bal)^2 + (\Delta\Bal)^2])^2}
\end{equation}
Equations~\ref{mugam} and \ref{errgam} were used to obtain $\mu_0f(k)$ and its esd  from
from the $\gamma$ values using $\sqrt{(\Delta\Bal^2+ \Delta\Bal^2)} =0.07$~\AA$^{-2}$ and $\Delta\bsm = 0.23$~fm.
Except for a few high angle reflections ($k^2$ large) the major contributor to the standard deviation is the
esd of the asymmetry itself.

Using the dipole approximation (Equation~\ref{dipole}) $\mu_0f(k)$ can be modelled using two parameters
$a_{j0}$ and $a_{j2}$:
\begin{equation}\label{sandl}
\mu_0f(k)=a_{j0}\jzerok + a_{j2} \jtwok
\end{equation}
where \jzerok and \jtwok are the values calculated by Blume et al. \cite{blume:62} for 
\smion.
\begin{table}
\caption{Parameters obtained by fitting equation~\ref{sandl} to the aligned magnetic moment
$\mu_0f(k)$ of Sm in \smgd\ measured at temperatures T }
\begin{center}
\begin{tabular}{rlllr}
\hline
T (K)& \multicolumn{1}{c}{a$_{j0}$}& \multicolumn{1}{c}{a$_{j2}$}& $\mu_0$(\mub)&N$_{\textrm{\scriptsize{obs}}}$\\
\hline
\multicolumn {4}{c}{Crystal X1}\\
 32.86 &$  -2.57(9)$&$\msp   2.78(8)$&$\msp   0.22(2)$&  15\\
 56.48 &$ -2.17(13)$&$\msp  2.29(13)$&$\msp 0.120(12)$&   3\\
 58.32 &$ -2.21(13)$&$\msp  2.34(13)$&$\msp   0.13(2)$&   4\\
 59.60 &$   -2.0(4)$&$\msp    2.1(4)$&$\msp  0.07(13)$&   4\\
 60.68 &$  -0.58(6)$&$\msp   0.68(6)$&$\msp   0.10(2)$&  17\\
 62.51 &$  -0.06(4)$&$\msp   0.11(4)$&$\msp  0.049(8)$&  13\\
 65.33 &$  -0.04(7)$&$\msp   0.08(7)$&$\msp  0.041(7)$&   4\\
 68.92 &$\msp   0.44(6)$&$  -0.42(6)$&$\msp 0.025(12)$&   4\\
 70.08 &$\msp    1.2(2)$&$   -1.3(2)$&$  -0.02(2)$&   3\\
 71.45 &$\msp   1.72(6)$&$  -1.79(6)$&$-0.068(10)$&  15\\
104.08 &$\msp    1.1(2)$&$   -1.2(2)$&$  -0.04(2)$&  14\\
\multicolumn {4}{c}{Crystal X2}\\
 32.95 &$  -1.70(6)$&$\msp   1.93(6)$&$\msp  0.235(8)$&  12\\
 49.17 &$   -1.8(3)$&$\msp    2.0(3)$&$\msp   0.18(3)$&   6\\
 58.92 &$  -1.44(7)$&$\msp   1.62(7)$&$\msp 0.179(10)$&   6\\
 62.93 &$  -1.49(6)$&$\msp   1.64(6)$&$\msp  0.149(6)$&   6\\
 64.80 &$  -1.41(5)$&$\msp   1.56(5)$&$\msp  0.149(5)$&   6\\
 67.03 &$ -0.40(14)$&$\msp  0.37(14)$&$  -0.02(4)$&  12\\
 68.68 &$\msp  1.32(11)$&$ -1.37(11)$&$-0.044(13)$&   8\\
 69.70 &$\msp  1.41(12)$&$ -1.47(12)$&$-0.056(14)$&   8\\
 73.44 &$\msp   1.34(5)$&$  -1.40(5)$&$ -0.059(7)$&  12\\

\hline
\end{tabular}
\end{center}
\label{results}
\end{table}%

The  $\mu_0f(k)$ values obtained from the measured asymmetries were sorted in order of their measurement temperatures and divided into groups within
which the temperature varied by no more than 1\degree~K. For each group containing more than two reflections
the values $a_{j0}$ and $a_{j2}$ and their estimated standard deviations were determined by a least squares fit to equation~\ref{sandl}. The values obtained are listed in \tabl{results}. 

\vspace{2ex}

\begin{figure}[h]
\begin{center}
\resizebox{0.75\textwidth}{!}{\includegraphics{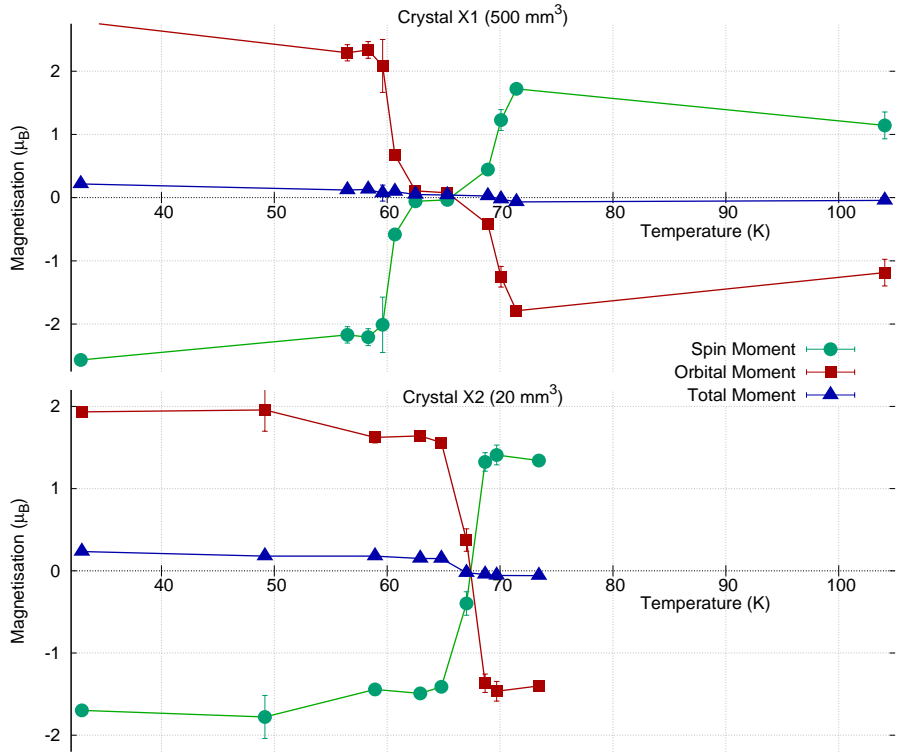}}
\caption{Variation with temperature of the spin and orbital contributions to the magnetization aligned by a field of 8T in \smgd.}
\label{magpars}
\end{center}
\end{figure}

The dipole approximation equates the parameters $a_{j0}$ and $a_{j2}$ with the spin and orbital components
of the magnetization and their sum to the ion's magnetization $\mu_0$. Figure~\ref{magpars} illustrates the variation of these parameters between 30 and 100 K. Below 60K both the spin and orbital aligned moments are about 
2~\mub\ with the orbital magnetization being slightly greater than the spin one. Between 60 and 65 K both fall
rapidly and change sign at \Tcomp\ $\approx 66$K (X1), $\approx 67$K (X2) above which the aligned moments recover but with the orbital moment again slightly greater than the spin one. If $M_L$ and $M_S$ are the orbital and spin magnetic moments of the
\smion\ ion: for $T<\Tcomp$ : $|M_L|>|M_S|$, at $T=\Tcomp\ : M_L=M_S=0$ and for $T>\Tcomp\ : |M_L|>|M_S|$ again.  

\section{Discussion}
Our results for the temperature variation of the spin and orbital moments of the \smion\ ion in \smgdx\ $x=0.024$ are very similar to those of Taylor et al. \cite{taylor02} for $x=0.018$  measured by non-resonant X-ray diffraction, but the precision 
of the neutron results is much better ($\pm\approx 0.1$ \mub)  compared with ($\pm\approx 0.8$ \mub) for the X-ray data.
Both sets of measurements show that the absolute value of the orbital magnetic moment of the \smion ion is always greater than that of its spin moment except at \Tcomp\ where both are zero. They also agree in indicating a reversal of both components with respect to the magnetizing field at \Tcomp\ which results in the ionic magnetization being opposed to  the magnetizing field at temperatures above \Tcomp. This unexpected result was attributed by Taylor et al.\cite{taylor02} to a combination of unwanted
beam movements from the the synchrotron bending magnet, and temperature fluctuations in the cryostat on 
reversing the applied field.  However neither of these fluctuations  perturb  the neutron measurements which are made with a stationary crystal under 
stable conditions of both temperature and field. In fact  the reversal of the apparent magnetization with respect to the magnetizing  field (neutron polarization direction) is a quite significant effect which  can be seen even in the raw aymmetry data: the low angle reflections 220 and 113  have significant intensity asymmetries which are positive at 30K and negative at 100K. It is however significant  that both experiments showing the magnetization reversal
derive the total moments $\mu_0$ from diffraction data with $k>0$, These do not include any contribution to the magnetization from conduction electrons for which the form factor is zero unless $k\approx0$. The apparent magnetization reversal at \Tcomp\ is therefore probably due to the presence of magnetized conduction electrons ferromagnetically coupled to the
 rare earth ion's spin moment.  At low temperature the difference  $\mu_0$ 
 between the orbital and spin components of the \smion\ is large and the magnetization direction is that of the dominant orbital part.
  As the temperature is raised $\mu_0$ falls and at \Tcomp\ is exceeded by the magnetization of the conduction electrons. It is then
 the total spin moment, the sum of the ionic spin moment and that of the conduction electrons, which is aligned parallel to the magnetizing field. The ionic magnetization, still dominated by its orbital component, is therefore aligned in the reverse direction.
 
\vspace{2ex}
  
  The observation that the reduction and reversal of both the spin and orbital components of the ionic magnetization takes place gradually over a range of about 10\degree~K, 
rather than abruptly at \Tcomp\ suggests that exact compensation occurs at different temperatures in different parts of the crystal. Since doping of \smal\ with Gd$^{3+}$ ions is needed to achieve ZMF it seems likely that the spin of the conduction electrons couples to the large spin moments of Gd$^{3+}$ ions. The random positioning of the doping ions within the \smal\ lattice
leads to a range in the conduction electron magnetization at the Sm sites and hence to  the range in \Tcomp.
Extrapolation of the absolute 
values of the spin and orbital moments across the transition region allows an estimation of their full values at \Tcomp.  The resulting difference
 0.15(2) \mub/Sm gives a value  for the conduction electron polarization necessary to obtain the ZMF state at \Tcomp.

\vspace{2ex}

In conclusion the present investigation confirms that the compensation phenomenon in the ZMF compound \smgdx\  is driven by the different temperature dependencies of the spin and orbital moments of the \smion\ ion.  We have obtained accurate values for the variation of both
the spin and orbital contributions to the magnetization of the \smion\ ion in a field of 8T.  These show that
the spin component of the ionic magnetic moment never exceeds the orbital one so that the ZMF state is not reached simply by equalization of the spin and orbital moments of \smion\ ions. It is already known that  doping of \smal\ with Gd is necessary to achieve the ZMF state. The gradual reversal of both the spin and orbital moments of \smion\ ions over a range of
$\approx$ 10\degree~K around \Tcomp\ suggests that compensation is achieved by enhancement of the conduction electron polarization by
the randomly substituted Gd$^{3+}$ doping ions.

\section{Acknowledgements}
We thank H. Adachi for providing us one (X2) of the two crystals used in the experiments.

\end{document}